\documentclass[amsmath, amssymb, aps, prb, twocolumn, notitlepage, longbibliography, superscriptaddress,eqsecnum]{revtex4-2}

\usepackage{graphicx}
\usepackage{calc}
\usepackage{cancel}
\usepackage{braket}
\usepackage{slashed}
\usepackage{wasysym}
\usepackage{dsfont}
\usepackage{float} 
\usepackage{amsthm,amsmath,amsfonts,amssymb,verbatim,color}
\usepackage{lipsum}
\usepackage{bm}
\usepackage{epsfig,slashed}
\usepackage{hyperref}
\usepackage{flafter}
\usepackage{booktabs}
\usepackage{makecell} 
\usepackage{bbm} 
\usepackage[export]{adjustbox} 
\usepackage{diagbox} 
\usepackage{makecell} 

\usepackage{xcolor}

\newcommand{\br}{{\bf r}}

\newcommand{\hs}{{\hat s}}

\def\hS{\mathds{S}}

\newcommand{\sgn}{\mathrm{sign}}

\begin{document}	
\title{Quasispins of vacancy defects and their interactions in disordered antiferromagnets}


	\author{Muhammad Sedik}
	\affiliation{Physics Department, University of California, Santa Cruz, California 95064, USA}

        \author{Shijun Sun}
	\affiliation{Physics Department, University of California, Santa Cruz, California 95064, USA}

	\author{Arthur P. Ramirez}
	\affiliation{Physics Department, University of California, Santa Cruz, California 95064, USA}
	
	\author{Sergey Syzranov}
	\affiliation{Physics Department, University of California, Santa Cruz, California 95064, USA}

\begin{abstract}
Vacancy defects in disordered
magnetic materials are known to act as effective spins, ``quasispins'', in response to an external magnetic field. In the dilute limit,
the contributions of such ``quasispins''
to the magnetic susceptibility $\chi_\text{vac}(T)\propto
N_\text{vac}/T$ are singular in the limit of low temperatures $T$
and match those of free spins. With increasing the density of vacancies, their interactions may become
essential.
{Motivated by frustrated and quasi-one-dimensional magnetic materials, we show by exact computation that the correlation of the quasispins of different vacancies match the spin-spin correlators in the vacancy-free materials for Ising chains with nearest-neighbour and next-to-nearest-neighbour interactions. We also compute the first virial correction to the susceptibility of a magnetic material 
due to the interactions of vacancy quasispins. We then provide a phenomenological generalisation of our results to a generic system that has short-range antiferromagnetic order and lacks long-range order. We predict that if the vacancy defect does not disrupt the short-range antiferromagnetic order around it,
the quasispin value
matches the value of spins of the magnetic atoms in the material, and the correlators of the quasispins of 
different vacancies match the spin-spin correlators in the vacancy-free material.}
\end{abstract}

	
\maketitle

\section{Introduction}

Often, quenched disorder in magnetic materials 
is seen as a hindrance that masks quantum-coherent effects.
For example, disorder effects may mimic the signatures of
the widely sought quantum spin liquid (QSL) state or replace it with the spin-glass state. 

On the other hand, the intentional introduction of impurities  
can serve as a powerful tool for uncovering the physics of clean materials. For example, the dependence of
the spin-glass-transition temperature on the concentration of vacancy defects reveals 
the existence of the ``hidden energy scale'', the property of the clean material which governs the low-temperature physics~\cite{SyzranovRamirez}. 
The effect of quenched disorder is particularly profound in materials that lack magnetic order, such as candidate materials for QSLs.

To date,
the most common form of quenched disorder in geometrically frustrated magnets, the largest class of candidate materials for QSLs, is vacancy defects. 
Such defects are typically created during synthesis by the controlled introduction of non-magnetic atoms
that substitute a fraction of the magnetic atoms of the material.
For example, in the compounds $SrCr_{9x}Ga_{12-9x}O_{19}$, known as 
SCGO, $Cr^{3+}$ spin-$3/2$ ions form a ``kagome bilayer'' at $x=1$ but can be replaced
by randomly located non-magnetic $Ga^{3+}$ with a tunable concentration~\cite{Ramirez:SCGO,Martinez:SCGO},
and in the compounds $Ni_{1-x}Zn_xGa_2S_4$, the spin-$1$ $Ni^{2+}$ ions can 
be substituted by the randomly located non-magnetic $Zn^{2+}$ atoms~\cite{NambuMaeno:NiGaS} (see Refs.~\cite{SyzranovRamirez} and \cite{RamirezSyzranov:review} for an extensive review of frustrated 
compounds with artificially introduced vacancy defects).

In order to use the vacancy defects
as a tool for investigating the properties of magnetic materials, 
it is essential to understand the effect of such vacancies on the
magnetic susceptibility. 
Although a non-magnetic atom that creates a vacancy defect has zero spin, it affects the collective behaviour of the surrounding spins.

It has been established~\cite{SchifferDaruka,LaForge} that
geometrically frustrated materials with vacancy defects
display a Curie-like contribution to the magnetic susceptibility given by
\begin{align}
    \chi_\text{vac}(T)=\frac{\langle \hS^2\rangle }{T}N_\text{vac},
    \label{QuasispinResponseGeneric}
\end{align}
where $N_\text{vac}$ is the number of vacancies
and $T$ is the temperature. This contribution
is equivalent to the magnetic susceptibility
of effective spins $\hS$, ``quasispins'', associated with the vacancies.

Another effect of vacancies in magnetic materials is the dilution of the magnetic atoms, 
which decreases the contribution
$\chi_\text{bulk}(T)$ of such atoms away from the vacancies. 
As the bulk susceptibility is normally either a non-singular or a less singular function of temperature 
than $\chi_\text{vac}\propto 1/T$, it is readily separated from the vacancy contribution $\chi_\text{vac}(T)$ in experiment.

The quasispin behaviour of vacancies should be expected in a generic magnetically-disordered material in which the magnetisation along a certain direction $z$ is a good quantum number, as follows from the fluctuation-dissipation theorem for the susceptibility $\chi_{zz}=\langle \hat M_z^2\rangle/T$, where $\hat M_z$ is the magnetisation of a correlated region of the material in which the vacancy is located.

The values of quasispins and their response to 
magnetic field have been the subject of research efforts in various systems, from 2D Heisenberg and classical antiferromagnets~\cite{SandvikDagottoScalapino,SachdevBuragonhainVojta,Sushkov,HoglundSandvik03,HoglundSandvik07,WollnyFritzVojta,WollnyVojta:vacancies,WollnyFritzVojta,WollnyVojta:vacancies,MaryasinZhitomirsky,Maryasin_2015} to spin chains and ladders~\cite{SandvikDagottoScalapino,SunRamirezSyzranov:1Dquasispin,KatsuraTsujiyama,Wortis1974,Bogani,GoupalovMattis,ValkovShustin}. 
Recently, we demonstrated~\cite{SunRamirezSyzranov:1Dquasispin} that quasispins generically emerge in systems as simple as 1D Ising spin chains with nearest-neighbour (NN) and
next-to-nearest-neighbour (NNN) interactions.

Despite extensive studies, a systematic understanding of the values of quasispins in a generic magnetically-disordered system remains elusive. Another unsettled question is the role of interactions between quasispins.
While the studies to date have focussed on the case of dilute vacancy defects, the
interactions between them may become important with increasing the defect density. 
This naturally raises the question about the possibility of novel types of magnetic order and/or quasispin-glass phases that emerge due to quasispin-quasispin interactions. 

Numerical simulations of the antiferromagnetic Heisenberg model with vacancies 
on a triangular lattice suggest~\cite{Maryasin_2015} that 
quasispin-quasispin correlations can enhance their collective susceptibility.
Furthermore,
understanding the collective behaviour of vacancy defects in frustrated magnetic materials may reveal the nature of the ``hidden energy scale'' in these systems~\cite{SyzranovRamirez}, the characteristic energy scale 
in clean frustrated magnets that determines the spin-glass-freezing temperature when vacancies are introduced.

In this paper, we study quasispins and their correlations in a magnetically-disordered system, 
motivated by both frustrated and quasi-one-dimensional magnets.
We argue
that in systems where the vacancy does not change the short-range magnetic order at low temperatures, the values of quasispins and quasispin-quasispin correlators match
the values of spins and spin-spin correlators in vacancy-free materials. 
We also compute the leading virial correction to the magnetic susceptibility that comes from the quasispin-quasispin correlations. 
We verify our conclusions by exact calculations of quasispins and quasispin correlators in the
Ising chain with NN and NNN interactions.
Our results can be tested experimentally in 
frustrated and quasi-one-dimensional magnetic materials.

The paper is organised as follows. 
In Section~\ref{sec:IsingDetails}, we exactly compute the magnetic susceptibilities
and quasispin-quasispin interactions in Ising chains with vacancy defects.
In Sec.~\ref{Sec:PhenomenologicalArgument}, we provide a phenomenological 
generalisation of our results to higher-dimensional systems that lack 
long-range magnetic order.
We discuss and summarise our results in Sec.~\ref{sec:conclusion}.







\begin{figure}[ht!]
	\centering
	\includegraphics[width = 3.3in]{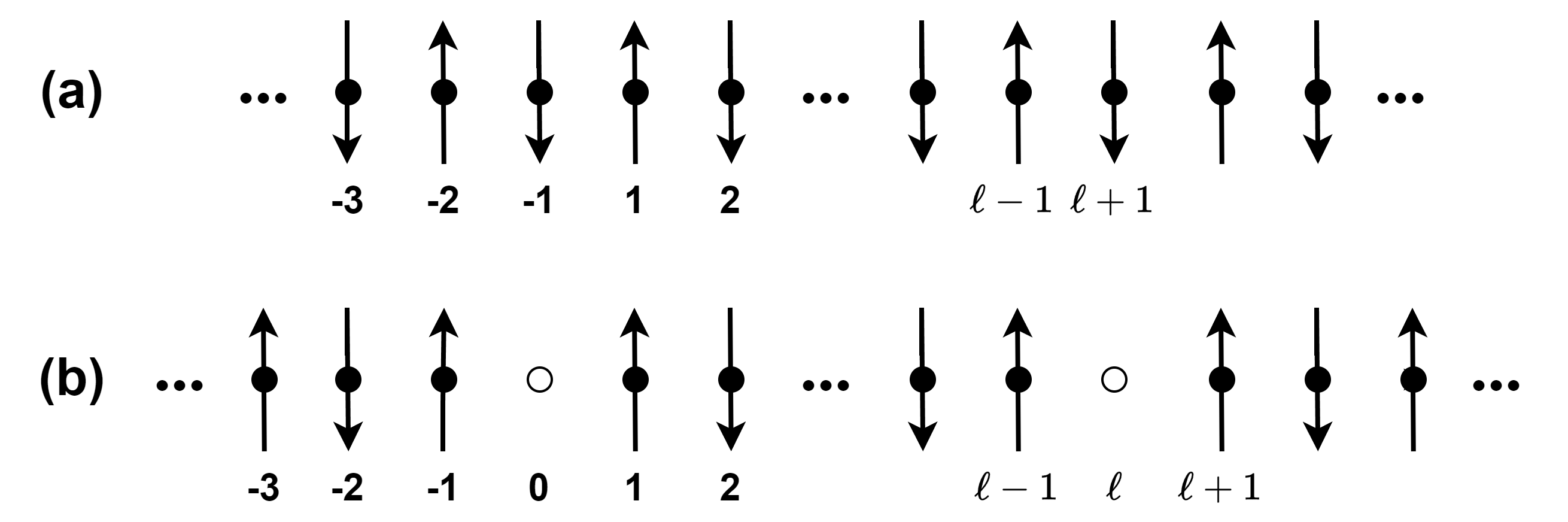}
	\caption{Spin configurations in the ground states of Ising chains with antiferromagnetic NN interactions and sufficiently weaker ferromagnetic NNN interactions (a) in the absence of vacancies and
   (b) in the presence of two vacancies distance $\ell$ apart.}
	\label{fig:l_geq_4}
\end{figure}


\section{Magnetic susceptibility in Ising chains: exact results}

\label{sec:IsingDetails}

In this section, we compute the magnetic susceptibility in Ising chains with 
vacancy defects. Such 1D systems allow for exact analytical treatment 
and give insights in the physics of quasispins in more complicated,
higher-dimensional systems.
We consider chains 
with AFM NN and sufficiently weaker NNN interactions, i.e. NNN interactions that do not alter the (AFM) ground state of the vacancy-free chain.

As we have shown in an earlier work~\cite{SunRamirezSyzranov:1Dquasispin},
vacancies in such chains have nonzero quasispins, $\sqrt{\hS^2}=1$,
if the NNN interactions are ferromagnetic (FM). In the case of AFM NNN interactions, 
the quasispins vanish, $\sqrt{\hS^2}=0$.
In the limit of dilute vacancy defects, their contributions to the 
magnetic susceptibility are described by Eq.~\ref{QuasispinResponseGeneric}.
With increasing the density of defects, quasispin-quasispin interactions become essential.
The purpose of this work is to investigate the effect of such interactions and 
their contribution to the magnetic susceptibility of the system.


\subsection{Generic expressions for susceptibility}


To investigate quasispin-quasispin interactions,
we consider a chain with two vacancies, located at sites $0$ and $\ell$, as shown
in Fig.~\ref{fig:l_geq_4}b, and compare its susceptibility to that of a chain without vacancies, shown in Fig.~\ref{fig:l_geq_4}a.

To compute the magnetic susceptibility of an Ising spin chain, we employ the fluctuation-dissipation theorem in the form
\begin{equation}
	{\chi}(T) = \frac{\langle {M}^2\rangle}{T}=\frac{1}{T}\sum_{i,j}\langle\sigma_{i}\sigma_{j}\rangle,
	\label{Eq:chi-correlation}
\end{equation}
where $\sigma_i=\pm 1$ is the spin at the $i$-th site of the chain, and 
$M=\sum_i \sigma_i$ is the total magnetisation of the system. Here, we have used that the average magnetisation is zero for Ising chains without external fields.

The spin correlation functions are computed using the same method as in our earlier work~\cite{SunRamirezSyzranov:1Dquasispin}. We outline the steps of the calculation in this section
and provide the details of the calculations in Appendix~\ref{Sec:two_vacancies}.

The difference between the susceptibilities $\chi(T)$ and $\chi_{0}(T)$ for the chains with and without vacancies can be expressed as
\begin{align}
    \chi(T) - \chi_0(T) =
    & \frac{4}{T}\sum_{i=2}^{\ell-2}\sum_{j=\ell-1}^{\infty}\left(\langle\sigma_i\sigma_j\rangle-\langle\sigma_i\sigma_j\rangle_{0}\right) \nonumber\\
    &+\frac{4}{T}\sum_{j=\ell+1}^{\infty} \left(\langle\sigma_{\ell-1}\sigma_{j}\rangle -\langle\sigma_{\ell-1}\sigma_{j}\rangle_{0}\right)\nonumber\\
    &+\frac{4}{T}\sum_{j=\ell+2}^{\infty} \left(\langle\sigma_{\ell+1}\sigma_{j}\rangle-\langle\sigma_{\ell+1}\sigma_{j}\rangle_{0}\right)\nonumber\\
    &+\frac{2}{T}\sum_{i=-\infty}^{1}\sum_{j=\ell-1}^{\infty}\left(\langle\sigma_i\sigma_j\rangle-\langle\sigma_i\sigma_j\rangle_{0}\right).
	\label{eq:differenceofsusceptibility}
\end{align}
Here, the correlators $\langle\cdots\rangle$ and $\langle\cdots\rangle_{0}$ are for chains with and without vacancies, respectively. In each of the sums, we extend the upper limits to infinity. This is a good approximation so long as the locations of the vacancies are sufficiently far away from the chain ends, i.e., the distance significantly exceeds the correlation length $\xi$.


\subsection{Quasispins and their correlations}





\subsubsection{AFM NN, FM NNN interactions}

In this subsection, we consider Ising chains with AFM NN and FM NNN interactions.
Utilising Eq.~\eqref{eq:differenceofsusceptibility} and computing the correlators $\langle\sigma_i\sigma_j\rangle$
and $\langle\sigma_i\sigma_j\rangle_0$ (see Appendix~\ref{Sec:two_vacancies} for details), we obtain 
the difference in the magnetic susceptibilities of chains with and without the two vacancies for the distances $|\ell|\geq2$ between the two vacancies:
\begin{align}
    \chi-\chi_{0}&\approx\frac{2}{T}\left[1+(-1)^\ell e^{-|\ell|/{\xi}}\right]\nonumber\\
    &-2\left[1+2(-1)^{\ell} e^{-|\ell|/{\xi}}\right] T^{-1}e^{-2 |J_2|/T}.
    \label{eq:susceptibility_dgt2}
\end{align}
The omitted corrections to the above results are of order $T^{-1}e^{-2J_{1}/T}$ or smaller.

For $|\ell|=1$, i.e. for the vacancies neighbouring each other, the two vacancies break the chain into two independent semi-infinite chains. The susceptibility of such a system consists of the ``bulk'' susceptibility~\cite{PiniRettori07}~\cite{SunRamirezSyzranov:1Dquasispin} and the susceptibility of two 
free ends of semi-infinite Ising chain~\cite{KassanOgly:OnedimensionalIM,SunRamirezSyzranov:1Dquasispin} 
\begin{equation}
    \chi \approx \frac{1}{2T} + \chi_{0}+\mathcal{O}\left[T^{-1}e^{-(2J_{1}+2|J_{2}|)/T}\right].
    \label{eq:susceptibility_d1}
\end{equation}

The {susceptibility}~\eqref{eq:susceptibility_dgt2} of a chain with two vacancies distance $|\ell|>1$ apart can be recast in the form 
\begin{align}
    \chi(T)= 
    \frac{2}{T}+\frac{2\langle \hS_1 \hS_2\rangle}{T}
    +\chi_\text{bulk}(T),
    \label{QuasispinCorrelationDefinition}
\end{align}
where the first term describes the contribution of two isolated quasispins of magnitude one; 
$\chi_\text{bulk}(T)\approx\chi_{0}(T)-2T^{-1}e^{-2|J_{2}|/T} = \left[1-\frac{2}{N}b(T)\right]\chi_{0}(T)$ is the contribution of the
bulk, and
\begin{equation}
    \left< \hS_1 \hS_{2}\right>=
    \left(-1\right)^\ell \exp\left({-\frac{|\ell|}{\xi}}\right) + \mathcal{O}\left(T^{-1}e^{-2|J_{2}|/T}\right)
    \label{eq:correlator_d}
\end{equation}
mimics the correlator of the quasispins of the two vacancies.
The quasispin-quasispin correlator given by Eq.~\eqref{eq:correlator_d}
matches the correlator of the spins in a defect-free chain.

If two vacancies are located at neighbouring sites ($|\ell|=1$),
we find, similarly to the case of $|\ell|>1$,
their quasispin correlation function to be given by
\begin{equation}
    \left< \hS_1 \hS_2\right>=
    -\frac{3}{4} + \mathcal{O}\left(T^{-1}e^{-2|J_{2}|/T}\right).
    \label{eq:correlator_1}
\end{equation}


\subsubsection{AFM NN, AFM NNN interactions}

{In Ising chains with AFM ground states and AFM NNN interactions, 
an isolated vacancy disrupts the short-range order, as the ground states in the presence and in the absence of a vacancy differ by flipping all spins to one side of the vacancy.
As a result, the values of quasispins, their correlators and the susceptibility are distinct from those in
chains with FM NNN interactions, in which the short-range order is not disrupted by vacancies.}

For AFM NNN interactions, vacancies have zero quasispins~\cite{SunRamirezSyzranov:1Dquasispin},
$\sqrt{\hS^2}=0$, i.e. do not exhibit singular $\propto 1/T$ contributions to the magnetic susceptibility $\chi(T)$. 
In this paper, we show (see Appendix~\ref{Sec:two_vacancies})
that the singular $\propto 1/T$ is also absent in a chain with
two vacancies, which signals the vanishing quasispin-quasispin
correlators,
\begin{equation}
    \langle\hS_{1}\hS_{2}\rangle = 0,
\end{equation}
for $\ell\geq2$, consistent with the vanishing of isolated quasispins.


\subsection{Virial corrections to the susceptibility}

\label{sec:virial}

{\it First virial correction to the susceptibility.} The magnetic susceptibility of an Ising chain with 
multiple vacancies is given by 
\begin{equation}
    \chi(T) = \frac{N_{\text{vac}}}{T} +  
    \frac{1}{T}\sum_{\substack{\alpha\neq\beta,\\\alpha,\beta=1}}^{N_{\text{vac}}} 
    \left<{\langle \hS_{\alpha}\hS_\beta\rangle}\right>_\text{loc}
    +\chi_\text{bulk}(T),
    \label{VirialStart}
\end{equation}
where
$\langle \hS_\alpha \hS_\beta\rangle$ is the correlator of quasispins 
$\alpha$ and $\beta$ and
$\langle\ldots\rangle_{\text{loc}}=\frac{1}{N^{N_{\text{vac}}}}\sum_{x_{1}=1}^{N}\sum_{x_{2}=1}^{N}\cdots\sum_{x_{N_{\text{vac}}}=1}^{N}\ldots$ is 
averaging with respect to the locations $x_1$, $x_2$, ..., $x_{N_{\text{vac}}}$ of the vacancies.
The first term in the right-hand-side of Eq.~\eqref{VirialStart} 
describes the contributions of 
individual quasispins to the susceptibility $\chi(T)$; the second term accounts for the correlations between
quasispins, and the last term is given by Eq.~\eqref{ChiBulk} and
describes the contribution of the bulk spins, which is independent of 
the locations of a given number of vacancies.

In the limit of dilute vacancies, $N_\text{vac}\ll N/\xi$, almost all vacancies may be considered isolated,
and only a small vacancy fraction, of order $\sim N_\text{vac}\xi/N$, 
are located within the correlation
length $\xi$ of other impurities. In what follows, we compute the first virial correction to the susceptibility
of isolated impurities, i.e. the correction that accounts for the probability of two vacancies being close to each other, while neglecting the possibility of clusters of three or more vacancies close to each other. In this approximation, the susceptibility~\eqref{VirialStart} can be rewritten as
\begin{equation}
    \chi(T) \approx \frac{N_{\text{vac}}}{T} +  
    \frac{1}{T}\frac{N_\text{vac}^2}{N^2}\sum_{x_1=1}^N \sum_{x_2=1}^N
    \langle \hS_1 \hS_2\rangle    +\chi_\text{bulk}(T),
    \label{Virial2}
\end{equation}
where $\hS_{1}$ and $\hS_{2}$ are the quasispins of two vacancies that may be located close to each other,
and the effect of the other vacancies on the correlator of $\hS_{1}$ and $\hS_{2}$ can be neglected.
In Eq.~\eqref{Virial2}, we have used that there are $N_{\text{vac}}\left(N_{\text{vac}}-1\right)/{2}\approx N_{\text{vac}}^{2}/2$ pairs of vacancies for $N_{\text{vac}}\gg1$.

Since the correlator
$\langle \hS_1 \hS_2\rangle $ depends only on the
distance $\ell=|x_{1}-x_{2}|$ between the vacancies, Eq.~\eqref{Virial2} can be simplified as
\begin{equation}
    \chi(T) \approx \frac{N_{\text{vac}}}{T} + 2\frac{N_{\text{vac}}^{2}}{NT} \sum_{\ell=1}^{\infty}  \langle \hS_{1}\hS_{2}\rangle
    +\chi_\text{bulk}(T).
    \label{eq:sparsevacancy}
\end{equation}

Substituting the values of the correlators~(\ref{eq:correlator_d}-\ref{eq:correlator_1}) 
into Eq.~\eqref{eq:sparsevacancy} gives the magnetic susceptibility~\eqref{Eq:QuasispinVirialCorrection} that includes the first virial correction
(see Appendix~\ref{Sec:virial_accurate}
for the details of the derivation of the viral correction).

\subsubsection{AFM NN, FM NNN interactions}

Such correlations between quasispins lead to corrections to the susceptibility $\chi(T)$ that are
non-linear in the number of impurities $N_\text{vac}$.
For Ising chains with ferromagnetic NNN interactions, we find
\begin{align}
    \chi(T)=\frac{N_\text{vac}}{T}
    \left\{1
    -\frac{1}{2}
    \frac{N_\text{vac}}{N}+
    {\cal O}
    \left[\left(\frac{N_\text{vac}}{N}\right)^2\right]
    \right\}
    +\chi_\text{bulk}(T),
    \label{Eq:QuasispinVirialCorrection}
\end{align}
where $N$ is the number of spins in the vacancy-free chain, and
\begin{align}
    \chi_\text{bulk}(T)=
    \frac{N-b(T)N_\text{vac}}{N}\chi_0(T)
    \label{ChiBulk}
\end{align}
describes the contribution of the ``bulk'' spins away from the vacancy; 
$\chi_0(T)\approx NT^{-1}e^{-(2J_1-4J_2)/T}$ 
is the susceptibility of the vacancy-free chain and $b(T)\approx e^{(2J_{1}-2J_{2})/T}$ is the effective ``size'' of a vacancy defect~\cite{SunRamirezSyzranov:1Dquasispin}, where $J_1>0$ and $J_2<0$ are the 
NN and NNN couplings.

The Curie-like contribution $N_\text{vac}/T$ in Eq.~\eqref{Eq:QuasispinVirialCorrection} is the contribution of isolated quasispins found in Ref.~\cite{SunRamirezSyzranov:1Dquasispin}. The second term in the curly brackets
gives the first virial correction to the system's susceptibility, i.e. the leading-order correction due to the quasispin-quasispin interactions.

The coefficient $-1/2$ in the first virial correction of the susceptibility~\eqref{Eq:QuasispinVirialCorrection}
of an Ising chain deviates from the respective coefficient in Eq.~\eqref{eq:GenericQuasispinBehaviour}
because the latter is derived under the assumption that vacancies do not disrupt the AFM order at short distances, and this assumption is violated in 1D by the configurations with two neighbouring vacancies, which break the chain into two independent half chains and thus break the AFM order. 
Indeed, the quasispin-quasispin correlator~\eqref{eq:correlator_d}
in a chain with two vacancies deviates from the spin-spin correlator in 
a vacancy-free chain if the vacancies are next to each other [cf. Eq.~\eqref{eq:correlator_1}], which leads to the deviation of the viral coefficient for a 1D chain with NN and NNN interactions from the generic prediction~\eqref{eq:GenericQuasispinBehaviour}.
In higher-dimensional materials, however, the virial correction will, in general, be described by
Eq.~\eqref{eq:GenericQuasispinBehaviour}.

\subsubsection{AFM NN, AFM NNN interactions}

{In Ising chains with AFM ground states and AFM NNN interactions, 
an isolated vacancy disrupts the short-range order, as the ground states in the presence and in the absence of a vacancy differ by flipping all spins to one side of the vacancy.
As a result, the values of quasispins, their correlators and the susceptibility are distinct from those in
chains with FM NNN interactions, in which the short-range order is not disrupted by vacancies.}

For AFM NNN interactions, vacancies have zero quasispins~\cite{SunRamirezSyzranov:1Dquasispin}, i.e. do not exhibit singular $\propto 1/T$ contributions to the magnetic susceptibility $\chi(T)$. We find in this paper that the quasispin correlation function vanishes
\begin{equation}
    \langle\hS_{1}\hS_{2}\rangle = 0 
\end{equation}
for $\ell\geq2$, in accordance with the vanishing of the quasispin. 

However, two vacancies may also occur at neighbouring sites, in which case they split the system into two independent parts to the left and to the right of the pair of vacancies.
The free ends of these parts give rise to a finite quasispin-like contribution $\chi^\prime(T)\propto 1/T$
to the magnetic susceptibility. As a result, the magnetic susceptibility averaged with respect to the
locations of vacancies also acquires a part with the Curie-type temperature dependence:
\begin{equation}
    \chi(T) = \frac{N_{\text{vac}}^{2}}{NT} + {\cal O}
    \left[\left(\frac{N_\text{vac}}{N}\right)^2\right] + \chi_{\text{bulk}}(T).
    \label{AFM-NN-FM-NNNsusceptiblity}
\end{equation}
Unlike the case of Ising chains with FM NNN interactions, the $\propto 1/T$ contribution comes entirely from the virial correction, hence the quadratic dependence $\propto N^2_\text{vac}$ on the number of vacancies.


\section{Correlations between quasispins
and virial corrections in a generic frustrated system}

In this section, we generalise the results we obtained for the Ising chains to more generic systems. While a system of quantum spins in higher dimensions cannot, in general, be described exactly analytically, we 
develop a set of phenomenological arguments for the values of quasispins and their correlators in a broad class of systems without long-range magnetic order and obtain the magnetic susceptibility to 
the first virial correction.

In what immediately follows, we consider a generic system 
in which the AFM order persists over a large correlation length $\xi$
at low temperatures,
but the fluctuations destroy the magnetic order at longer distances.
We will assume also that the presence of the vacancy does not disrupt the short-range 
magnetic order near the vacancy at low temperatures. Short-range magnetic order unchanged by a vacancy should generically 
be expected in 
in (quasi-)2D materials with Heisenberg spins on bipartatite lattices.
Sufficiently strong NN interactions
in such materials ensure short-range order, while long-range order
with a continuous order parameter is forbidden by the Mermin-Wagner theorem at $T>0$. A similar trend could be expected in 2D weakly frustrated systems.
Another example of a system in which a vacancy does not affect
the short-range order is given by the considered above
Ising chains with AFM NN and FM NNN interactions.
Short-range order may also be robust against impurities in QSLs
on frustrating lattices. However, 
obtaining exact microscopic states of such QSLs presents a significant challenge and requires further theoretical advance.
The results for the magnetic susceptibility obtained in this paper may serve as a verification of the respective effect (or lack thereof) of vacancy defects on the short-range order.

We assume also that the magnetization $\hat M_z$ in the system we consider is a good quantum number.
This corresponds to the majority of 
magnetic materials with strong exchange spin-spin interactions,
which may be presumed to be isotropic.

Due to the commutation of the magnetization $\hat M_z$ and the Hamiltonian of the system, the fluctuation-dissipation relation 
$\chi_{zz}=\left<\hat M_z^2\right>/T$ can be used for obtaining the
magnetic susceptibility along the $z$ axis.


\subsection{Qualitative argument for the quasispin values 
and quasispin-quasispin correlations}

\label{Sec:PhenomenologicalArgument}

{\it The value of the quasispin of an isolated vacancy.}
The configurations of spins that contribute to low-temperature observables are ordered antiferromagnetically near the vacancy, which makes it convenient to define the ``missing spin'' $\hat s_z$
at the location of the vacancy as the spin that would complete the local AFM spin texture.
The magnetisation $\hat m=\hat M_z-\hat s_z$ of the system with a vacancy along the $z$ direction can be understood as the difference between the magnetisation $M_z$ of a vacancy-free chain and one
missing spin $s_z$.
In what immediately follows, we neglect
the correlator $\langle \hat{M}_z \hs_z\rangle$
averaged over the states of the system with a vacancy and make the approximation
\begin{align}
    \left<(\hat M_z- \hat s_z)^2\right>\approx \left<\hat M_z^2\right>+\left<\hat s_z^2\right>.
    \label{eq:neglectCorr}
\end{align}
The suppression of the correlator $\langle \hat M_z \hat s_z\rangle$, which we use here,
can be understood as follows. Under the made assumption of the vacancy not altering the short-range AFM order, 
$\langle \hat M_z \hat s_z\rangle\approx\langle \hat M_z \hat s_z\rangle_0$,
where $\left<\ldots\right>_0$ is our convention for 
averaging over the state of the vacancy-free system.
Summing this relation over all possible
locations of the vacancy and using the fluctuation-dissipation theorem gives the estimate
$\langle \hat M_z \hat s_z\rangle\approx\langle \hat M_z \hat s_z\rangle_0=\langle \hat M_z^2\rangle_0/N\equiv TN^{-1}\chi_{zz}^0(T)$ which vanishes in the limit of low temperatures $T\rightarrow 0$.
Indeed, for higher-dimensional frustrated systems, this vanishing follows from the Curie-Weiss behaviour~\cite{Ramirez:ReviewFrustrated}
$\chi_{zz}^0(T)\propto \left(T+|\theta_W|\right)^{-1}$
of the magnetic susceptibility of the vacancy-free system. 
For 1D Ising chains, the correlator $\langle \hat M_z \hat s_z\rangle$ 
gives rise to the ``vacancy-size'' contribution in Eq.~\eqref{ChiBulk}, exponentially suppressed, $\propto e^{-2|J_2|/T}$, in the limit of low temperatures.
Qualitatively, the suppression of the correlator  $\langle \hat M_z \hat s_z\rangle$
comes from the 
AFM order in the vicinity of the vacancy, which results in the suppressed contribution of the magnetisation of the spins around the vacancy to $M_z$.

Neglecting, under the made assumption, the difference between the quantities $\langle \hs_z^2\rangle$
and $\langle \hs_z^2\rangle_0$ and utilising 
the fluctuation-dissipation relation 
$\chi(T)=\left<(\hat{M}_z-\hs_z)^2\right>/T$, Eq.~\eqref{eq:neglectCorr} leads to 
the total magnetic susceptibility 
\begin{align}
    \chi_{zz}(T)=\frac{\left<\hs_z^2\right>}{T}N_\text{vac}+
    \chi_\text{bulk}^0(T),
    \label{PhenomenologicalChi}
\end{align}
in a system with dilute vacancy defects (i.e. in the limit 
$N_\text{vac}\rightarrow 0$), where $\chi_\text{bulk}^0(T)$ is the susceptibility of the vacancy-free system.
Equation~\eqref{PhenomenologicalChi} shows that the fluctuations of the quasispins along
the $z$ axis match those of the spins in the vacancy-free system.

Equation~\eqref{PhenomenologicalChi} is consistent with the results for the 1D Ising chains that we obtained in Sec.~\eqref{sec:IsingDetails}. Neglecting the correlations between
the quasispin and the bulk spins, which are suppressed by the factor 
$e^{-2|J_2|/T}\ll 1$, Eq.~\eqref{PhenomenologicalChi} matches the exact result,
Eqs.~\eqref{Eq:QuasispinVirialCorrection} and \eqref{ChiBulk} for the Ising
chain with AFM NN and FM NNN interaction up to the ${\cal O}\left(N_\text{vac}\right)$
contributions.


{\it Two-quasispin correlations.}
Similarly, if two vacancies are introduced in the system, and these vacancies do not alter
the short-range AFM order, one can make the approximation
\begin{align}
    &\left<(\hat M_z-\hat s_{1z}-\hat s_{2z})^2\right> \nonumber\\
    &\approx\left<\hat M_z^2\right>
    +\left<\hat s_{1z}^2\right>+\left<\hat s_{2z}^2\right>+2\left<\hat s_{1z}\hat s_{2z}\right>.
    \label{TwoVacanciesPhenomenology}
\end{align}
Because the susceptibility in a system with two vacancies is given by $\chi_{zz}(T)=\left<(\hat M_z-\hat s_{1z}-\hat s_{2z})^2\right>/T$, provided the magnetisation along the $z$ axis is a good quantum number, the first term in the right-hand side of Eq.~\eqref{TwoVacanciesPhenomenology} corresponds to the bulk-spin contribution, and the other terms describe the quasispins contribution $\left<\left(\hS_{1z}+\hS_{2z}\right)^2\right>$. 
Therefore, the quasispin-quasispin correlators match the spin-spin correlators in the vacancy-free system,
\begin{align}
    \left< \hS_{1z}\hS_{2z}\right>\approx
    \left< \hat s_{1z}\hat s_{2z}\right>_0.
    \label{CorrelatorGeneric}
\end{align}
This conclusion is consistent with the results of our exact calculations for Ising spin chains.

We use the correlators~\eqref{CorrelatorGeneric} to compute the virial corrections to the susceptibility, i.e. the leading-order corrections that account for pairwise correlations between quasispins, in a generic frustrated system in Sec.~\eqref{sec:virial}.


\subsection{Virial correction in a generic system}

Using the results of the phenomenological arguments of the previous section,
we make a prediction for the quasispin behaviour in higher-dimensional systems that 
lack magnetic order at low temperatures and may exhibit quantum behaviour 
of the spins (Heisenberg spins).
As discussed in Sec.~\ref{Sec:PhenomenologicalArgument}, we expect the quasispin-quasispin correlator in systems where vacancies do not affect short-range magnetic order to match the spin-spin correlator in the vacancy-free system. That allows us to obtain the susceptibility including the first virial correction in the form
\begin{align}
        \chi(T) \approx \frac{N_{\text{vac}}}{T}\left<\hs_{\br z}^2\right>_0
        + \frac{N_{\text{vac}}^{2}}{N^{2} T} \sum_{\substack{\br,\\\br^\prime\neq\br}} \left<\hs_{\br z}\hs_{\br^\prime z}\right>_0
    +\chi_\text{bulk}(T),
    \label{SusceptibilityGenericPre}
\end{align}
where $\left<\hs_{\br z}\hs_{\br^\prime z}\right>_0$ is the correlator of the $z$-components of spins at sites $\br$ and $\br^\prime$ in a vacancy-free system.

Introducing the total magnetisation $M_z=\sum_\br s_{\br z}$ in the $z$ direction, the susceptibility \eqref{SusceptibilityGenericPre} can be rewritten as
\begin{align}
        \chi(T) \approx 
        &\frac{N_{\text{vac}}}{T}\left<\hs_{\br z}^2\right>_0
        + \frac{N_{\text{vac}}^{2}}{N^2 T} \left< \hat{M}_z^2\right>_0
        \nonumber\\
        &-\frac{N_{\text{vac}}^{2}}{N T}  \left<\hs_{\br z}^2\right>_0
    +\chi_\text{bulk}(T),
    \label{ChiGeneric2}
\end{align}
where $\hat s_{\br z}$ is the $z$-component of a spin at an arbitrary site $\br$ in the vacancy-free system.

In systems with long correlation lengths $\xi$ significantly exceeding the lattice spacing, spins may be considered ordered antiferromagnetically on length scales $r\ll\xi$, with a suppressed magnetisation.
For such systems, $\langle\hat{M}_z^2\rangle\equiv N\langle\hat{M}_z \hat s_z\rangle\ll N\langle \hs_z^2\rangle$, in accordance with the argument in Sec.~\ref{Sec:PhenomenologicalArgument},
and the second term in the right-hand-side of Eq.~\eqref{ChiGeneric2} can be neglected, leading to a magnetic susceptibility in the form
\begin{align}
    \chi_{zz}(T)=
    \frac{N_\text{vac}}{T}\left<\hs_z^2\right>_0
    \left\{
    1-\frac{N_\text{vac}}{N}+
    {\cal O}
    \left[\left(\frac{N_\text{vac}}{N}\right)^2\right]
    \right\}
    \nonumber\\
    +\chi_\text{bulk}(T).
    \label{eq:GenericQuasispinBehaviour}
\end{align}

The first term in the curly brackets in Eq.~\eqref{eq:GenericQuasispinBehaviour}
describes the contribution of isolated quasispins to the magnetic susceptibility in the limit of small density
of the vacancies. 
The contribution of a single isolated vacancy is given by $\left<\hs_z^2\right>_0/T$
and matches the contribution of a free spin with the same variance $\left<\hs_z^2\right>_0$ of the $z$-component
as that of a spin in a vacancy-free material.
Equation~\eqref{eq:GenericQuasispinBehaviour} indicates, in particular, that in an isotropic 
Heisenberg model, the value of the quasispin of a vacancy matches the bulk spin in the defect-free material.
Furthermore, we show that the quasispin correlation function of two vacancies matches the spin
correlation function in a vacancy-free material.

The second term in the curly brackets in Eq.~\eqref{eq:GenericQuasispinBehaviour} describes the leading virial correction to the magnetic susceptibility that comes from pairwise interactions between the quasispins.
The ${\cal O}
    \left[\left(\frac{N_\text{vac}}{N}\right)^2\right]$ contribution in Eq.~\eqref{eq:GenericQuasispinBehaviour}
are the virial terms of higher orders in $N_\text{vac}$, which we do not compute in this paper and leave for future studies.


\section{Discussion and conclusion}

\label{sec:conclusion}

We have studied the effect of vacancy defects on the magnetic susceptibility of materials that lack long-range magnetic order at lower temperatures, exemplified by geometrically frustrated magnets.
The main effect of such vacancies is the emergence of quasispins, 
effective magnetic moments associated with the defects.

In the literature, there have been discussed several other mechanisms of vacancy defects' contributions to the magnetic susceptibility. 
Apart from quasispins associated with individual vacancies,
the Curie-like contribution has been proposed to come from
``orphan spins''~\cite{SchifferDaruka},
i.e. free spins of magnetic atoms disconnected from the bulk of the spins by the vacancies.
This scenario is, however, at odds numerically with the measured value of the Curie constant and the vacancy density as inferred from stoichiometry ensured by synthesis~\cite{LaForge}. 

Another proposed effect of vacancies is the emergence of
``half-orphan'' spins~\cite{Henley:HalfOrphans},
which are not completely disconnected but are weakly linked to the rest of the system, e.g. due to being next to a pair of vacancies.
Half-orphan spins have been theoretically shown~\cite{SenDamleMoessner,PatilDamle:HalfOrphans} to display the susceptibility of free fractional spins. However, the concentration of
``half-orphan'' spins is small and will scale nonlinearly ($\propto N_\text{vac}^Z/T$, where $Z>1$ is the number of vacancies near the quasispin) with the vacancy density $N_\text{vac}$ in the limit of small $N_\text{vac}$.
Such a contribution is suppressed in comparison with the leading quasispsin
contribution~\eqref{QuasispinResponseGeneric} to the susceptibility.
Quasispins are, therefore, the dominant effect of the vacancy defects on the magnetic response of the material.

We argue that if a vacancy defect does not alter 
the short-range antiferromagnetic order around a vacancy, the value of the quasispin matches the value of the spins of the magnetic atoms in the material.

When the vacancy defects are dilute, the quasispins act as isolated spins.
With increasing vacancy concentration, quasispin-quasispin interactions can become important.
If vacancies do not alter the short-range antiferromagnetic order, the
quasispin-quasispin correlation function matches the spin-spin correlator in the vacancy-free system.
For such materials, we also evaluate the first virial correction to the susceptibility 
that comes from pairwise quasispin-quasispin correlations.

We verify our predictions byexact analytical calculations of the magnetic susceptibility for Ising spin chains with nearest-neighbour (NN) and next-to-nearest-neighbour (NNN) interactions, whose ground states are antiferromagnetic (AFM). For chains with AFM NN and sufficiently weaker FM NNN interactions,
an isolated vacancy does not destroy the short-range AFM order. As we found previously~\cite{SunRamirezSyzranov:1Dquasispin},
such chains have vacancy quasispins $S=1$, in accordance with the generic arguments given in this paper.
In this work, we also find that the quasispin-quasispin correlators in such chains match the spin-spin correlators in vacancy-free chains, in agreement with the generic arguments, with the exception 
of a pair of vacancies located next to each other, effectively breaking the chain into two semi-infinite chains.
We have explicitly computed such correlators and the magnetic susceptibility of the chains, including the first virial correction, Eq.~\eqref{Eq:QuasispinVirialCorrection}:
\begin{align}
    \chi(T)=\frac{N_\text{vac}}{T}
    \left\{1
    -\frac{1}{2}
    \frac{N_\text{vac}}{N}+
    {\cal O}
    \left[\left(\frac{N_\text{vac}}{N}\right)^2\right]
    \right\}
    +\chi_\text{bulk}(T),
    \label{QuasispinVirialCorrectionFinal}
\end{align}
where $N$ and $N_\text{vac}$ are the numbers
of sites and vacancies, respectively; and $\chi_\text{bulk}$ is
the appropriately rescaled contribution of the bulk spins.

For chains with AFM NN and FM NNN interactions, the quasispin value vanishes~\cite{SunRamirezSyzranov:1Dquasispin}, $S=0$. In this case,
however, vacancy defects still contribute to the magnetic susceptility
due to the configurations with two defects located on neighbouring sites.
Such chains display the
Curie-like $\propto N_\text{vac}^2/T$ vacancy susceptibility
given by Eq.~\eqref{AFM-NN-FM-NNNsusceptiblity}.



We qualitatively generalise our arguments for the values of quasispins and their interactions to systems in higher dimensions and predict a magnetic susceptibility in the form
\begin{align}
    \chi_{zz}(T)=
    \frac{N_\text{vac}}{T}\left<\hs_z^2\right>_0
    \left\{
    1-\frac{N_\text{vac}}{N}+
    {\cal O}
    \left[\left(\frac{N_\text{vac}}{N}\right)^2\right]
    \right\}
    \nonumber\\
    +\chi_\text{bulk}(T),
\end{align}
where $\left<\hs_z^2\right>_0$ is the variance of a spin in 
a vacancy-free system, and the first and the second terms in the curly
brackets describe, respectively, 
the contributions of dilute vacancy defects and the first virial correction.

Our conclusions can be experimentally verified
in quasi-1D materials with antiferromagnetic ground states, such as BCVO~\cite{Zhao:BCVO,Niesen:BCVO,Zou:BCVO}, as well as higher-dimensional geometrically frustrated systems.

	
\section{Acknowledgements}

This work has been supported by
the NSF grant DMR-2218130.

\bibliography{references}


	\onecolumngrid
	\vspace{2cm}
	
	\cleardoublepage

	
	\setcounter{equation}{0}
	\setcounter{figure}{0}
	\setcounter{enumiv}{0}

	\appendix

 \section{Derivation of the quasispin-quasispin
 correlators in Ising chains}
 \label{Sec:two_vacancies}
 

In this appendix, we compute the 
correlators of quasispins of two vacancy defects
in an Ising chain with NN and NNN interactions, characterised, respectively,
by the couplings $J_1$ and $J_2$. We focus on chains with antiferromagnetic grounds states
corresponding to 
\begin{align}
    J_1>2|J_2|. 
\end{align}
We compute the correlators $\langle\sigma_i\sigma_j\rangle$ of spins in such Ising 
chains with and without vacancies and utilise
Eq.~(\ref{Eq:chi-correlation}) to find the difference $\chi(T)-\chi_0(T)$ of their magnetic susceptibilities.

Correlations between spins in chains with and without vacancies
can be conveniently computed by mapping spin configurations to a gas of domain walls~\cite{SunRamirezSyzranov:1Dquasispin}, where a domain wall is an excitation consisting in simultaneously flipping all the spins to the right of a given one. The energy of creating a domain wall in an antiferromagnetically ordered defect-free chain is given by 
\begin{align}
    E_D=2J_1-4J_2>0.
\end{align}
In the limit of low temperatures, $T\ll E_D$, the domain-wall excitations are very dilute. A vacancy-free chain in this limit presents a sequence of antiferromagnetic domains separated by domain walls.

The presence of vacancies modifies the energies $E_i$ of domain walls between sites $i$ and $i+1$
that neighbour vacancies. In the limit of dilute domain walls we consider, i.e. at low temperatures $T\ll E_i$, 
the domain walls are very sparse, and configurations of spins with two or more domain walls located on neighbouring links of the chain can be neglected. 
In this approximation, the correlator of two spins $\sigma_i$ and $\sigma_j$ with $j>i$ can be approximated as
\begin{align}
    \langle \sigma_i\sigma_j\rangle\approx(-1)^{j-i} 
    \frac{1-\sum_{{k_1}=i}^{j-1}e^{-\beta E_{k_1}}+\sum_{i\leq{k_1}<{k_2}<j-1}e^{-\beta(E_{k_1}+E_{k_2})}-\ldots
    }
    {1+\sum_{{k_1}=i}^{j-1}e^{-\beta E_{k_1}}+\sum_{i\leq{k_1}<{k_2}<j-1}e^{-\beta(E_{k_1}+E_{k_2})}+\ldots
    },
    \label{eq:CorrelatorPartitionFunction}
\end{align}
where $\beta = 1/T$.
The first term in the numerator in Eq.~(\ref{eq:CorrelatorPartitionFunction}) accounts for the contribution
of the AFM ordered state without domain walls to the correlator. The second and third terms in the numerator
describe the contribution of spin configurations with, respectively, two and three domain walls. 
Because adding a domain wall to a spin configuration changes the sign of the correlator, the signs in the numerator
in Eq.~\eqref{eq:CorrelatorPartitionFunction} are alternating.
The denominator gives the partition function.


Equation~(\ref{eq:CorrelatorPartitionFunction}) can be simplified by noticing
that the $n^{\text{th}}$ term in the numerator is the sum over all the possible products of $n$ distinct exponential $(e^{-\beta E_{k_1}},e^{-\beta E_{k_2}},\dots,e^{-\beta E_{k_n}})$ 
corresponding to $n$ out of the $(j-i)$ domain walls:  
\begin{align}
    \langle \sigma_i\sigma_j\rangle\approx(-1)^{j-i}\frac{\prod_{k=i}^{j-1}(1-e^{-\beta E_k})}{\prod_{k=i}^{j-1}(1+e^{-\beta E_k})}
    =\prod_{k=i}^{j-1}g(E_k),
    \label{eq:CorrelationWithg}
\end{align}
where we have introduced, for simplicity, the function
\begin{align}
 g(E_k)\approx -\sgn\left(E_k\right)\left(1-2e^{-\beta |E_k|}\right)+O\left(e^{-2\beta E_k}\right).
\label{eq:gDefinition}
\end{align}

In the presence of vacancies,
the AFM-ordered state is not necessarily the ground state of the chain. Some domain walls lead to spin configurations with lower energy and those domain walls have energy $E_k<0$. For example, a domain wall to the right of site $1$ in Fig.~\ref{fig:l_2}(b) has energy $E_1=2J_2$, which leads to a lower energy configuration if $J_2<0$ i.e. spins around the vacancy tend to align due to the ferromagnetic NNN interaction. 




\subsubsection*{Possible domain-wall energies}

There exist multiple energies $E_i$ of domain walls depending on the mutual locations
of vacancies. In what follows, we focus on the case of two vacancies in the chain.
There are four types of domain walls, whose energies in isolation are given by
\begin{align}
    E=2J_2, 2J_1-2J_2, 2J_1-4J_2, 2J_1.
\end{align}
An energy of $2J_2$ corresponds to the domain wall at the location of a vacancy.
A domain wall next to a vacancy has the energy $2J_1-2J_2$. If two vacancies are separated by two sites (as shown in Fig.~\ref{fig:l_3}), then a domain between these two sites has an energy of $2J_1$. All other domain walls, i.e. domain walls at least two sites away from vacancies, have the same energy $E_D=2J_1-4J_2>0$
as in the vacancy-free chain.



For further convenience, we introduce notations for the function $g(E_i)$ given by
Eq.~\eqref{eq:gDefinition} for each type of domain walls:
\begin{align}
  g(2J_2)\equiv g_1, \ g(2J_1-2J_2)\equiv g_2, \ g(E_D)\equiv g_3, \text{ and } g(2J_1)\equiv g_4.
    \label{eq:gisDefinition}
\end{align}

In what immediately follows, we consider spin chains with two vacancies for various distances $\ell$ between the vacancies (measured in the lattice spacing). As we clarify below, the cases $\ell=1,2,3$ require special consideration, while the correlators for $\ell \geq 4$ are given by a generic expression.


\subsection{The case $\ell=1$ (vacancies on neighbouring sites)}





\begin{figure}[h!]
    \centering
    \includegraphics[height=1.7in]{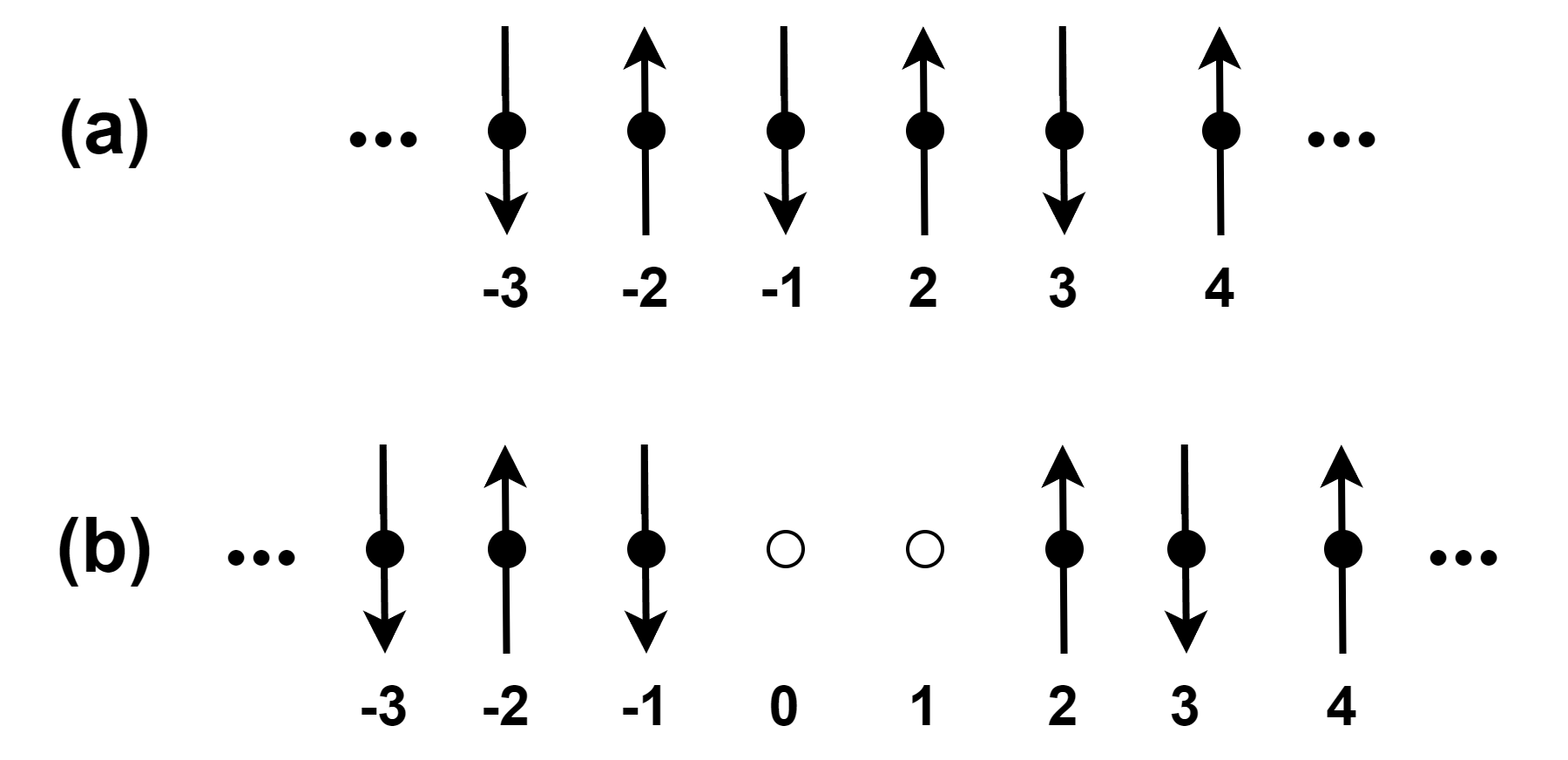}
    \caption{
    Spin configuration in the AFM-ordered state of an Ising chain with antiferromagnetic NN interactions $(J_1>0)$ and weaker NNN interactions. (a) is the chain with no vacancies, whereas (b) has two vacancies at site $0$ and $1$. There is no energy change if only spins with label $\geq 2$ are flipped.}
         \label{fig:l_1}
\end{figure}

In this subsection, we compute the difference $\chi(T)-\chi_0(T)$
of the magnetic susceptibility $\chi(T)$ of a chain with two vacancies on neighbouring sites (corresponding to the distance between vacancies $\ell=1$ in units of the lattice spacing)
and the susceptibility $\chi_0(T)$ of the vacancy-free chain. It is convenient to label the spins in the chain with the vacancies and in its vacancy-free counterpart as shown in 
Fig.~\ref{fig:l_1}.

In the absence of vacancies at low temperatures $T\ll J_1, E_D$,
the spin-spin correlators in a 1D chain are given by~\cite{stephenson1970two,robert1984decay}\cite{SunRamirezSyzranov:1Dquasispin}
\begin{align}
    \langle\sigma_i\sigma_j\rangle_0\approx\left\{\begin{array}{cc}
         (g_3)^{j-i-2},& i\leq -1, 2\leq j   \\
         (g_3)^{|j-i|},& i,j\leq -1 \text{ and } 2\leq i,j,
    \end{array}\right.
    \label{eq:Correlations0l1}
\end{align}
where the function $g_3$ is introduced in Eqs.~\eqref{eq:gDefinition} and (\ref{eq:gisDefinition}). To find spin-spin correlators in a chain with the vacancies, we note that two vacancies on neighbouring sites
split the chain into two independent half-chains.
The energy cost of a domain wall at a link adjacent to the end of a half-chain (between sits $-2$ and $-1$
or $2$ and $3$) is given by $2J_1-2J_2>0$, while the energy of a domain wall at any other location is $E_D=2J_1-4J_2$.
\begin{table}[h!]
    $$\begin{array}{|c||c|c|c|c|}
    \hline
         \langle\sigma_i\sigma_j\rangle& i\leq -2 & i=-1& i=2 &3 \leq i\\
         \hline\hline
         j \leq -2 &g_3^{|j-i|} & g_2g_3^{|j+2|} & 0 & 0\\
         \hline
         j =-1 &g_2g_3^{|i+2|} &1  & 0&0\\
         \hline
         j = 2 & 0& 0 &1 &g_2g_3^{i-3} \\
         \hline
         3 \leq j  &0& 0&g_2g_3^{j-3} & g_3^{|j-i|}\\
    \hline
    \end{array}$$
    \caption{
     The correlators $\langle\sigma_i\sigma_j \rangle$ 
     in a chain with vacancies on two neighbouring sites $0$ and $1$.
     The spins are labelled as shown in Fig.~\ref{fig:l_1}. The quantities $g_2$
     and $g_3$ are defined by Eqs.~\eqref{eq:gDefinition} and \eqref{eq:gisDefinition}.
    }
    \label{tab:Correlatorl1}
\end{table}
Using these values of domain-wall energies and Eq.~\eqref{eq:CorrelationWithg}, we obtain the spin-spin correlators 
$\langle\sigma_i\sigma_j\rangle$ at all locations in the chain, as summarised in Table~\ref{tab:Correlatorl1}.

The difference of the magnetic susceptibilities between the chain with two vacancies and the vacancy-free chain is given by
\begin{align}
    \frac{\chi-\chi_0 }{2\beta}&=\left(\sum_{i=-\infty}^{-2}\langle\sigma_i\sigma_{-1}\rangle-\langle\sigma_i\sigma_{-1}\rangle_0\right)+\left(\sum_{j=3}^{\infty}\langle\sigma_{2}\sigma_{j}\rangle-\langle\sigma_{2}\sigma_{j}\rangle_0\right)-\left(\sum_{i=-\infty}^{-1}\sum_{j=2}^{\infty}\langle\sigma_{i}\sigma_{j}\rangle_0\right).
    \label{eq:Chi-Correlationl1}
\end{align}
We note that the difference $\chi-\chi_0$ in Eq.~\eqref{eq:Chi-Correlationl1} comes from the correlations of the spins adjacent to the vacancies (at sites $-1$ and $2$) with the other spins in the chain. The correlators $\langle\sigma_i\sigma_j\rangle$ of spins not involving the spins at sites $-1$ and $2$ 
match the respective correlators in the vacancy-free chain.

Utilising Eq.~\eqref{eq:Chi-Correlationl1}, the correlators $\langle\sigma_i\sigma_j\rangle_0$
in the clean chain given by Eq.~\eqref{eq:Chi-Correlationl1} and the correlators in the chain with the vacancies summarised in Table~\ref{tab:Correlatorl1} gives
\begin{align}
    \chi-\chi_0\approx4\beta\left[g_2-g_3\right](1-g_3)^{-1}-2\beta g_3 (1-g_3)^{-2}.
     \label{eq:Chi-gisl1}
\end{align}
In the limit low temperatures, {$\beta (2J_1-2J_2) ,\beta E_D \gg 1$}, the values of $g_2$ and $g_3$ can be approximated as $g_3\approx g_2\approx -1$, which gives
\begin{align}
        \chi-\chi_0\approx\frac{\beta}{2}+O\left[e^{-2\beta (J_1-J_2)}\right].
     \label{eq:Chi-finall1}
\end{align}
In the case of two vacancies at neighbouring sites under consideration,
the magnetic-susceptibility difference $\chi-\chi_0$ can be viewed as coming from the
two free ends of the two spin half-chains. An open end of a half-chain 
contributes $\beta/4$ to magnetic susceptibility at low temperatures \cite{SunRamirezSyzranov:1Dquasispin}, hence, a two-end contribution
of $\beta/2$ in Eq.~\eqref{eq:Chi-finall1}.


\subsection{$\ell=2$ (Vacancies at sites $0$ and $2$)}

\begin{figure}[h!]
    \centering
    \includegraphics[height=1.7in]{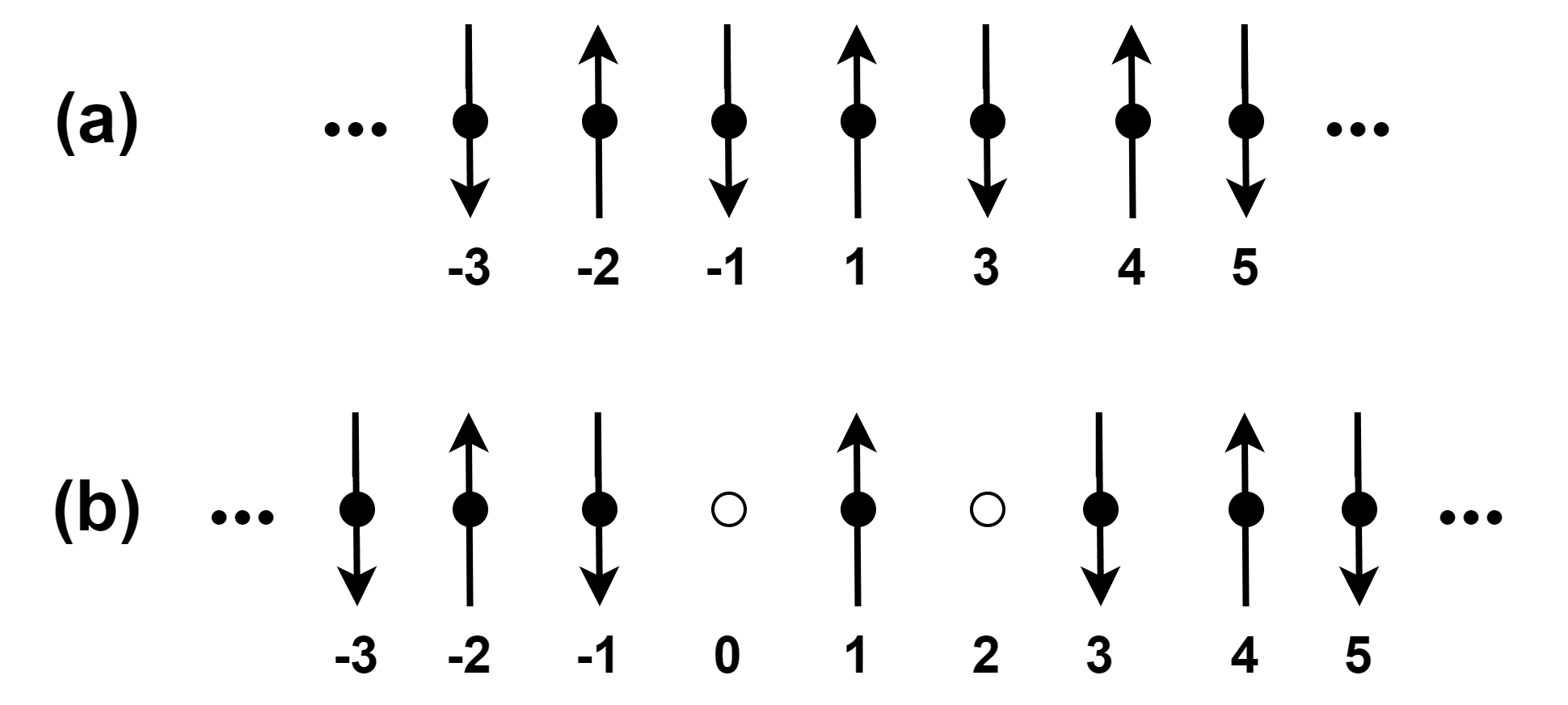}
    \caption{Spin configuration in the AFM-ordered state of an Ising chain with antiferromagnetic NN interactions $(J_1>0)$ and weaker NNN interactions in (a) a vacancy-free chain and in in (b) a chain with vacancies at sites $0$ and $2$.}
    \label{fig:l_2}
\end{figure}

In this subsection, we consider the case of two vacancies separated by one site.
We consider vacancies are at sites $0$ and $2$ as shown in Fig.~\ref{fig:l_2}.
A domain wall in the vacancy-free chain has an energy cost of $E_D$.
Therefore, the correlations between spins for this chain are given by 
\begin{align}
    \langle\sigma_i\sigma_j\rangle_0\approx\left\{\begin{array}{cc}
         (g_3)^{j-i-2},& i\leq -1, 3\leq j,   \\
         (g_3)^{|j-i|},& i,j\leq -1 \text{ and } 3\leq i,j, 
    \end{array}\right., \ \text{ and } \ \langle \sigma_i\sigma_1\rangle_0\approx(g_3)^{i-{2\sgn{(i)}}},
    \label{eq:Correlations0l2}
\end{align}
where $g_3$ is defined by Eqs.~\eqref{eq:gDefinition} and (\ref{eq:gisDefinition}).

For a chain with two vacancies, a domain-wall excitation between sites $-1(1)$ and $1(3)$ has an energy of $2J_2$; the energy cost of a domain-wall between sites $-1(3)$ and $-2(4)$ is $2J_1-2J_2$, and 
a domain wall at any other location has the energy $E_D=2J_1-4J_2$. 
Utilising these values of energies of domain walls and Eq.~(\ref{eq:CorrelationWithg}), we compute the correlators $\langle \sigma_i \sigma_j \rangle$ for all spins in the chain with vacancies. The values of the correlators are summarised in Table \ref{tab:Correlatorl2}.
    \begin{table}[h!]
    $$\begin{array}{|c||c|c|c|c|c|}
    \hline
         \langle\sigma_i\sigma_j\rangle& i\leq -2 & i=-1& i=1 &i=3 &4 \leq i\\
         \hline\hline
         j \leq -2 &g_3^{|j-i|} & g_2g_3^{|j+2|}& g_1g_2g_3^{|j+2|}&g_1^2g_2g_3^{|j+2|} &g_1^2g_2^2g_3^{i-j-6}\\
         \hline
         j =-1 &g_2g_3^{|i+2|} &1  & g_1&g_1^2 &g_1^2g_2g_3^{i-4}\\
         \hline
         j = 1 & g_1g_2g_3^{|i+2|}& g_1 &1 &g_1 &g_1g_2g_3^{i-4}\\
         \hline
         j=3   &g_1^2g_2g_3^{|i+2|} & g_1^2& g_1& 1&g_2g_3^{i-4}\\
         \hline
         4 \leq j  &g_1^2g_2^2g_3^{j-i-6}& g_1^2g_2g_3^{j-4}& g_1g_2g_3^{j-4}&g_2g_3^{j-4} & g_3^{|j-i|}\\
    \hline
    \end{array}$$
    \caption{
    The correlators $\langle\sigma_i\sigma_j \rangle$ in a spin chain with two vacancies,
    at sites $0$ and $2$, separated by one site. The sites are labeled as shown in
    Fig.~\ref{fig:l_2}. The quantities $g_1,g_2$ and $g_3$ are defined by Eqs.~(\ref{eq:gisDefinition}) and \eqref{eq:gDefinition}.}
    \label{tab:Correlatorl2}
\end{table}

Taking into account the symmetry relations $\langle \sigma_i \sigma_j\rangle=\langle \sigma_j \sigma_i\rangle$ and that the diagonal entries in Table \ref{tab:Correlatorl2} match the respective correlators in the vacancy-free chain for the spins labeled as shown in Fig.~\ref{fig:l_2},
the difference of the magnetic susceptibilities between the chain with vacancies and the vacancy-free chain is given by
\begin{align}
    \frac{\chi-\chi_0}{2\beta}=&\sum_{i=-\infty}^{-2}\sum_{j=4}^{\infty}\left(\langle \sigma_i\sigma_j\rangle-\langle \sigma_i\sigma_j\rangle_0\right)+2\sum_{j=4}^{\infty}\Big(\langle\sigma_{3}\sigma_{j}\rangle+\langle\sigma_{1}\sigma_{j}\rangle+\langle\sigma_{-1}\sigma_{j}\rangle-\langle\sigma_{3}\sigma_{j}\rangle_0-\langle\sigma_{1}\sigma_{j}\rangle_0-\langle\sigma_{-1}\sigma_{j}\rangle_0\Big) \notag \\ &+2\langle\sigma_{-1}\sigma_{1}\rangle+\langle\sigma_{-1}\sigma_{3}\rangle-2\langle\sigma_{-1}\sigma_{1}\rangle_0+\langle\sigma_{-1}\sigma_{3}\rangle_0.
    \label{eq:Chi-Correlationl2}
\end{align}
By substituting the correlators given by Eq.~(\ref{eq:Correlations0l2}) and Table~\ref{tab:Correlatorl2} into Eq.~(\ref{eq:Chi-Correlationl2}), we obtain the magnetic susceptibility difference in terms of $g_1,g_2$ and $g_3$ as 
\begin{align}
    \frac{\chi-\chi_0}{2\beta}\approx2g_1-2g_3+g_1^2-g_3^2+2\left(g_2+g_1g_2+g_1^2g_2-g_3-g_3^2-g_3^3\right)\left(1-g_3\right)^{-1}+\left(g_1^2g_2^2-g_3^4\right)\left(1-g_3\right)^{-2}.
    \label{eq:Chi-gisl2}
\end{align}
At low temperatures, $\beta |J_2|, \ \beta (2J_1-2J_2), \ \beta E_D \gg 1$, we approximate $g_3=g_2\approx -1$ and $g_1\approx \sgn(J_2)\left(-1+2e^{-2\beta |J_2|}\right)$, and
the magnetic-susceptibility difference is given by 
\begin{align}
   \chi-\chi_0\approx2\big[1-\sgn(J_2)\big]\beta+2\big[2\sgn(J_2)-1\big]\beta e^{-2\beta|J_2|}+O\left[\beta e^{-4\beta |J_2|},\beta e^{-2\beta (J_1-J_2)}\right].
    \label{eq:Chi-finall2}
\end{align}

\subsection{$\ell=3$ (Vacancies at sites $0$ and $3$)}

\begin{figure}[h!]
    \centering
    \includegraphics[height = 1.7in]{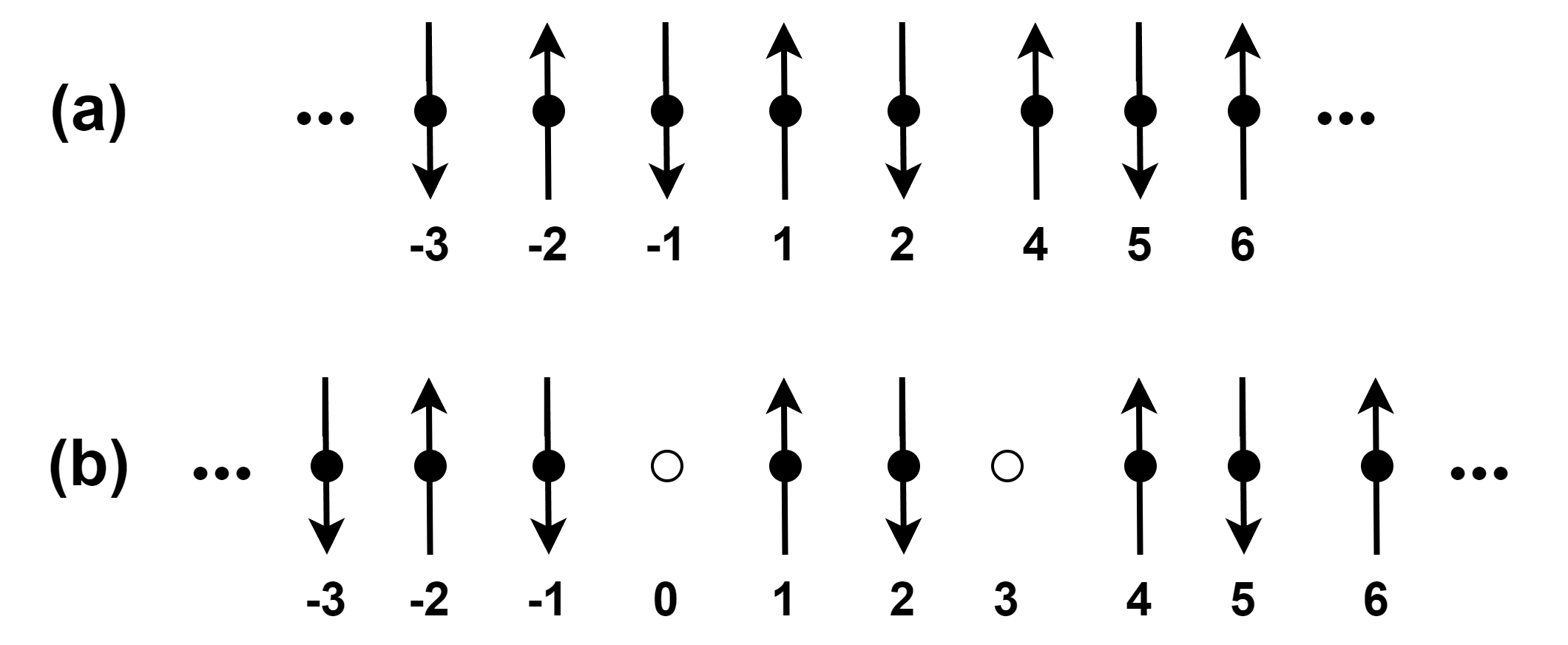}
    \caption{AFM ordered states of Ising chains with antiferromagnetic NN $(J_1>0)$ and weaker NNN interactions for (a) a vacancy-free chain and
    (b) a chain with vacancies at sites $0$ and $3$.}
    \label{fig:l_3}
\end{figure}

The case of vacancies separated by two sites also requires special consideration.
We consider two vacancies at sites $0$ and $3$, as shown in Fig.~\ref{fig:l_3}.
The correlations between spins for the vacancy-free chain are given by 
\begin{align}
    \langle\sigma_i\sigma_j\rangle_0\approx\left\{\begin{array}{cc}
         (g_3)^{j-i-2},& i\leq -1, 4\leq j,   \\
         (g_3)^{|j-i|},& i,j\leq -1 \text{ and } 4\leq i,j, 
    \end{array}\right., \ \langle \sigma_i\sigma_1\rangle_0\approx(g_3)^{i-{2\sgn{(i)}}}, \ \text{ and } \ \langle \sigma_i\sigma_2\rangle_0\approx(g_3)^{i-{3\sgn{(i)}}},
    \label{eq:Correlations0l3}
\end{align}
where $g_3$ is defined by Eqs.~(\ref{eq:gisDefinition}) and \eqref{eq:gDefinition}.

For a chain with two vacancies, the energy cost of a domain wall at the location of 
a vacancy, i.e. between sites $-1(2)$ and $1(4)$, is $2J_2$.
A domain wall between $-2(5)$ and $-1(4)$ has an energy of $2J_1-2J_2$.
The energy of a domain wall between sites $2$ and $3$ is $2J_1$.
The energy of a domain wall at any other location is given by $E_D$.
Taking into account these energies, we obtain the spin-spin correlators $\sigma_i\sigma_j$
summarised in Table~\ref{tab:Correlatorl3}.
    \begin{table}[h]
    $$\begin{array}{|c||c|c|c|c|c|c|}
    \hline
         \langle\sigma_i\sigma_j\rangle& i\leq -2 & i=-1& i=1 &i=2 &i=4 &5 \leq i\\
         \hline\hline
         j \leq -2 &g_3^{|j-i|} &g_2g_3^{|j+2|} &g_1g_2g_3^{|j+2|} &g_1g_2g_3^{|j+2}g_4 &g_1^2g_2g_3^{|j+2|}g_4 &g_1^2g_2^2g_3^{i-j-7}g_4\\
         \hline
         j =-1 &g_2g_3^{|i+2|} &1 &g_1 &g_1g_4 &g_1^2g_4 &g_1^2g_2g_3^{i-5}g_4\\
         \hline
         j = 1 & g_1g_2g_3^{|i+2|}& g_1 &1 &g_4 & g_1g_4 &g_1g_2g_3^{i-5}g_4\\
         \hline
         j=2   &g_1g_2g_3^{|i+2}g_4 & g_1g_4& g_4&1&g_1 &g_1g_2g_3^{i-5}\\
         \hline
         j=4   &g_1^2g_2g_3^{|i+2|}g_4& g_1^2g_4& g_1g_4&g_1 &1 &g_2g_3^{i-5}\\
         \hline
         5 \leq j  & g_1^2g_2^2g_3^{j-i-7}g_4& 
         g_1^2g_2g_3^{j-5}g_4& g_1g_2g_3^{j-5}g_4&g_1g_2g_3^{j-5} &g_2g_3^{j-5} & g_3^{|i-j|}\\
    \hline
    \end{array}$$
    \caption{The correlators $\langle\sigma_i\sigma_j \rangle$ in a chain with vacancies
    at sites $0$ and $3$. The spins are labelled as shown in Fig.~\ref{fig:l_3}. 
    The quantieis $g_1,g_2,g_3$ and $g_4$ are defined by Eqs.~\eqref{eq:gDefinition} and (\ref{eq:gisDefinition}).}
    \label{tab:Correlatorl3}
\end{table} 

The magnetic susceptibility 
can be represented in the form
\begin{align}
    \frac{\chi-\chi_0}{2\beta}&=\sum_{i=-\infty}^{-2}\sum_{j=5}^{\infty}\left(\langle \sigma_i\sigma_j\rangle-\langle \sigma_i\sigma_j\rangle_0\right)\notag\\ &+2\sum_{j=5}^{\infty}\Big(\langle\sigma_{4}\sigma_{j}\rangle+\langle\sigma_{2}\sigma_{j}\rangle+\langle\sigma_{1}\sigma_{j}\rangle+\langle\sigma_{-1}\sigma_{j}\rangle-\langle\sigma_{4}\sigma_{j}\rangle_0-\langle\sigma_{2}\sigma_{j}\rangle_0-\langle\sigma_{1}\sigma_{j}\rangle_0-\langle\sigma_{-1}\sigma_{j}\rangle_0 \Big)\notag\\ &+2\langle\sigma_{-1}\sigma_{1}\rangle+2\langle\sigma_{-1}\sigma_{2}\rangle+\langle\sigma_{-1}\sigma_{4}\rangle+\langle\sigma_{1}\sigma_{2}\rangle-2\langle\sigma_{-1}\sigma_{1}\rangle_0-2\langle\sigma_{-1}\sigma_{2}\rangle_0-\langle\sigma_{-1}\sigma_{4}\rangle_0-\langle\sigma_{1}\sigma_{2}\rangle_0.
    \label{eq:Chi-Correlationl3}
\end{align}
By substituting the correlators in Eq.~(\ref{eq:Correlations0l3}) and Table~\ref{tab:Correlatorl3} into Eq.~(\ref{eq:Chi-Correlationl3}), we obtain the magnetic susceptibility difference between the chains with and without vacancies:
\begin{align}
    \frac{\chi-\chi_0}{2\beta}\approx& 2g_1-2g_3+2g_4g_1-2g_3^2+g_4g_1^2-g_3^3+g_4-g_3\notag\\
    &+ 2\left(g_2+g_1g_2+g_4g_1g_2+g_4g_1^2g_2-g_3-g_3^2-g_3^3-g_3^4\right)\left(1-g_3\right)^{-1} +\left(g_4g_1^2g_2^2-g_3^5\right)\left(1-g_3\right)^{-2}.
    \label{eq:Chi-gisl3}
\end{align}
At low temperatures $\beta |J_2|, \ \beta(2J_1-2J_2), \ \beta E_D \gg 1$,
we can approximate $g_4=g_3=g_2\approx -1$, $g_1\approx \sgn(J_2)\left(-1+2e^{-2\beta |J_2|}\right)$ and
\begin{align}
    \chi-\chi_0 \approx 2\beta e^{-2\beta|J_2|}+O\left[\beta e^{-4\beta |J_2|},\beta e^{-2\beta J_1}\right].
     \label{eq:Chi-finall3}
\end{align}


\subsection{$\ell \geq 4$ (Vacancies at sites $0$ and $\ell\geq4$)}

In this subsection, we consider two vacancies at sites $0$ and $\ell \geq 4$, i.e. separated by at least three other sites.
We will label the spins in the chain with vacancies and in its vacancy-free counterpart as shown in Fig.~\ref{fig:l_geq_4}.
The spin-spin correlators in the vacancy-free chain are given by 
\begin{align}
   \langle\sigma_i\sigma_j\rangle_0\approx\left\{\begin{array}{cc}
         (g_3)^{j-i-1},& \quad \left(i\leq -1, 1\leq j\leq \ell-1\right)  \text{ and }\left(1\leq i\leq \ell-1, \ell+1\leq j  \right),\\
         (g_3)^{j-i-2},& \quad i\leq -1 \text{ and } \ell+1\leq j,\\
         (g_3)^{|j-i|},& \quad \left(i,j\leq -1 \right)\text{ and } \left(1\leq i,j\leq \ell-1\right)\text{ and } \left(\ell+1\leq i,j\right),
    \end{array}\right.
    \label{eq:Correlations0l_geq_4}
\end{align}
where $g_3$ is defined by Eqs.~\eqref{eq:gDefinition}and (\ref{eq:gisDefinition}). For the system with two vacancies, the energy cost of a domain wall at the location of a vacancy between sites $-1$ $(\ell-1)$ and $1$ $(\ell+1)$ is given by $2J_2$. The energy of a domain wall between sites $-2$ $(\ell-2)$ and $-1$ $(\ell-1)$ equals $2J_1-2J_2$ and matches the energy of a domain wall
between sites $1$ $(\ell+1)$ and $2$ $(\ell+2)$. Finally, the energy
of a domain wall at any other location is given by $E_D=2J_1-4J_2>0$. Similarly to previous actions, these energies allow
us to compute the spin-spin correlators in the system with 
vacancies, which are summarised in Table~\ref{tab:Correlatorl_geq_4}.
    \begin{table}[h]
    $$\begin{array}{|c||c|c|c|c|c|c|c|c|}
    \hline
         \langle\sigma_i\sigma_j\rangle& i\leq -2 & i=-1& i=1 & 2\leq i\leq \ell-2 & i=\ell-1 & i=\ell+1 &\ell+2 \leq i\\
         \hline\hline
         j\leq -2 & g_3^{|j-i|} & g_2g_3^{|j+2|}& g_1g_2g_3^{|j+2|} & g_1g_2^2g_3^{i-j-4} &g_1g_2^3g_3^{\ell-j-6}  &g_1^2g_2^3g_3^{\ell-j-6}  &g_1^2g_2^4g_3^{i-j-8}\\ 
         \hline
         j=-1 & g_2g_3^{|i+2|} &1& g_1 &g_1g_2g_3^{i-2}  &g_1g_2^2g_3^{\ell-4}  & g_1^2g_2^2g_3^{\ell-4} &g_1^2g_2^3g_3^{i-6}\\
         \hline
         j=1&g_1g_2g_3^{|i+2|}& g_1 & 1 &g_2g_3^{i-2}  &g_2^2g_3^{\ell-4}  &g-1g_2^2g_3^{\ell-4}  &g_1g_2^3g_3^{i-6}\\ 
         \hline
         2\leq j\leq \ell-2 &g_1g_2^2g_3^{j-i-4}&g_1g_2g_3^{j-2}&g_2g_3^{j-2} &g_3^{|j-i|} &  g_3^{\ell-j-2}& g_1g_2g_3^{\ell-j-2}  &g_1g_2^2g_3^{i-j-4}\\ 
         \hline
         j=\ell-1&g_1g_2^3g_3^{\ell-i-6}&g_1g_2^2g_3^{\ell-4}&g_2^2g_3^{\ell-4}  &g_3^{\ell-i-2}  & 1& g_1 &g_1g_2g_3^{i-\ell-2}\\ 
         \hline
         j=\ell+1&g_1^2g_2^3g_3^{\ell-i-6}&g_1^2g_2^2g_3^{\ell-4}&g_1g_2^2g_3^{\ell-4}  &g_1g_2g_3^{\ell-i-2} &  g_1& 1 &g_2g_3^{i-\ell-2}\\ 
         \hline
         \ell+2 \leq j&g_1^2g_2^4g_3^{j-i-8} & g_1^2g_2^3g_3^{j-6} &g_1g_2^3g_3^{j-6}  &g_1g_2^2g_3^{j-i-4}  & g_1g_2g_3^{j-\ell-2} & g_2g_3^{j-\ell-2}  &g_3^{|j-i|}\\ 
    \hline
    \end{array}$$\caption{The correlation $\langle\sigma_i\sigma_j \rangle$ in a chain with vacancies at sites $0$ and $\ell \geq 4$. The spins are labeled as shown in Fig.~\ref{fig:l_geq_4}. The quantities $g_1,g_2$ and $g_3$ are defined by Eqs.~\eqref{eq:gDefinition} and (\ref{eq:gisDefinition}).}
    \label{tab:Correlatorl_geq_4}
\end{table}

The difference of magnetic susceptibilities of the chains with and without vacancies is given by
\begin{align}
    \frac{\chi-\chi_0}{2\beta}&=
    \left(2\sum_{i=-\infty}^{-2}\sum_{j=2}^{\ell-2}+\sum_{i=-\infty}^{-2}\sum_{j=\ell+2}^{\infty}\right)\Big(\langle\sigma_i\sigma_j\rangle-\langle\sigma_i\sigma_j\rangle_0\Big)+2\sum_{j=\ell+2}^{\infty}\Big(\langle\sigma_{\ell+1}\sigma_{j}\rangle+\langle\sigma_{\ell-1}\sigma_{j}\rangle+\langle\sigma_{1}\sigma_{j}\rangle+\langle\sigma_{-1}\sigma_{j}\rangle\notag\\ 
    &-\langle\sigma_{\ell+1}\sigma_{j}\rangle_0-\langle\sigma_{\ell-1}\sigma_{j}\rangle_0-\langle\sigma_{1}\sigma_{j}\rangle_0-\langle\sigma_{-1}\sigma_{j}\rangle_0\Big)+2\sum_{j=2}^{\ell-2}\Big(\langle\sigma_{1}\sigma_{j}\rangle+\langle\sigma_{-1}\sigma_{j}\rangle-\langle\sigma_{1}\sigma_{j}\rangle_0-\langle\sigma_{-1}\sigma_{j}\rangle_0\Big)\notag\\ &+2\langle\sigma_{-1}\sigma_{1}\rangle+2\langle\sigma_{-1}\sigma_{\ell-1}\rangle+\langle\sigma_{1}\sigma_{\ell-1}\rangle+\langle\sigma_{-1}\sigma_{\ell+1}\rangle-2\langle\sigma_{-1}\sigma_{1}\rangle_0-2\langle\sigma_{-1}\sigma_{\ell-1}\rangle_0-\langle\sigma_{1}\sigma_{\ell-1}\rangle_0-\langle\sigma_{-1}\sigma_{\ell+1}\rangle_0.
    \label{eq:Chi-Correlationl_geq_4}
\end{align}
By substituting the correlators in Eq.~(\ref{eq:Correlations0l_geq_4}) and Table \ref{tab:Correlatorl_geq_4} into Eq.~(\ref{eq:Chi-Correlationl_geq_4}), we obtain the magnetic-susceptibility difference given by
\begin{align}
    \chi-\chi_0\approx 2\beta A + 2(-1)^{\ell}\beta \ B|g_3|^{\ell-4}, 
    \label{eq:Chi-gisl_geq_4}
\end{align}
where the $\ell$-independent quantities $A$ and $B$
are given by
\begin{subequations}
\label{eq:A_B_definitions}
    \begin{align}
        A(\beta)=&2(g_1-g_3)+4\left(g_2+g_1g_2-g_3-g_3^2\right)\left(1-g_3\right)^{-1}+2(g_1g_2^2-g_3^3)\left(1-g_3\right)^{-2}, \label{eq:A(beta)}\\
        B(\beta)=&g_2^2+2g_1g_2^2+g_1^2g_2^2-g_3^{2}-2g_3^{3}-g_3^{4}+2\left(g_1g_2^3+g_1^2g_2^3-g_3^4-g_3^{5}-g_2g_3-g_1g_2g_3+g_3^{2}+g_3^{3}\right)\left(1-g_3\right)^{-1}\notag\\
        &+2(-g_1g_2^2g_3+g_3^4+\frac{1}{2}g_1^2g_2^4-\frac{1}{2}g_3^{6})\left(1-g_3\right)^{-2}.
    \end{align}
\end{subequations}

At low temperatures, $\beta |J_2|, \ \beta (2J_1-2J_2), \ \beta E_D \gg 1$, one may approximate $g_3=g_2\approx -1$, $g_1\approx \sgn(J_2)\left(-1+2e^{-2\beta |J_2|}\right)$ and  
\begin{equation}
    A(\beta)\approx  \frac{1}{2}\big[1-\sgn(J_2)\big]+\sgn(J_2)e^{-2\beta |J_2|},\quad B(\beta)\approx \frac{1}{2}\big[1-\sgn(J_2)\big]+\big[\sgn(J_2)-1\big]e^{-2\beta |J_2|}.\label{eq:ABapprox}
\end{equation} 
Utilising the correlation length $\xi=-\left(\ln|g_{3}|\right)^{-1}$ for Ising chains with antiferromagnetic NN and {weaker} NNN interactions~\cite{SunRamirezSyzranov:1Dquasispin}, Eq.~\eqref{eq:Chi-gisl_geq_4} can be written as
\begin{align}
    &\chi-\chi_0\approx
    \nonumber \\
    &\beta\big[1-\sgn(J_2)\big]\left[1+(-1)^\ell e^{-\frac{\ell}{\xi}}\right]+2\left\{\sgn(J_2)+\big[\sgn(J_2)-1\big](-1)^\ell e^{-\frac{\ell}{\xi}}\right\}\beta e^{-2\beta |J_2|}+O\left[e^{-4\beta |J_2|},e^{-2\beta (J_1-J_2)}\right].
     \label{eq:Chi-finall_geq_4}
\end{align}

The susceptibility~\eqref{eq:Chi-finall_geq_4} is consistent with the 
susceptibility~\eqref{QuasispinCorrelationDefinition} expected based on
the quasispin model of the magnetic properties of vacancies. 
For $J_2>0$, Eq.~\eqref{eq:Chi-finall_geq_4} corresponds to the quasispin value
$\sqrt{\langle \hS^{2}\rangle} = 1$ and the ``vacancy size'' $b(T) = e^{(2J_{1}-2J_{2})/T}$.
The $\ell$-dependent contribution to Eq.~\eqref{eq:Chi-finall_geq_4} describes the quasispisn-quasispin
correlation function: $\langle \hS_1 \hS_{2}\rangle=B g_3^{\ell-4}$  (for $\ell \geq 4$).
For an arbitrary sign of the coupling $J_2$, the quasispin-quasispin correlators are summarised by the equations
\begin{subequations}\label{Aeq:correlator_accurate}
    \begin{align}
        \langle \hS_{1}\hS_{2}\rangle&
        \approx \frac{1}{4}\big[2 \ \sgn(J_2)-1\big]-\sgn(J_2)e^{-2\beta |J_2|}+O\left(e^{-4\beta |J_2|},e^{-2\beta J_1+2\beta J_2)}\right),\qquad\text{for} \quad \ell=1,\\
        \langle \hS_{1}\hS_{2}\rangle&
        \approx (-1)^\ell e^{-\frac{\ell}{\xi}}\left\{\frac{1}{2}\big[1-\sgn(J_2)\big]+\big[\sgn(J_2)-1\big]e^{-2\beta |J_2|}+O\left(e^{-4\beta |J_2|},e^{-2\beta J_1+ 2\beta J_2)}\right)\right\},  \qquad\text{for}\quad \ell\geq2.
    \end{align}
\end{subequations}
 

 \section{Magnetic susceptibility for chains with finite concentration of vacancies}
 \label{Sec:virial_accurate}

In this section, we provide the details of the calculation of the first virial correction to the magnetic susceptibility of an Ising chain with vacancies due to pairwise quasispin-quasispin correlations. 
We note first that, according to Eq.~\eqref{Aeq:correlator_accurate},
\begin{align}
    \sum_{\ell=1}^{\infty}  \langle \hS_{1}\hS_{2}\rangle &\approx \left\{\frac{1}{4}\big[2 \ \sgn(J_2)-1\big]-\sgn(J_2)e^{-2\beta |J_2|}\right\}+\left\{\frac{1}{2}\big[1-\sgn(J_2)\big]+\big[\sgn(J_2)-1\big]e^{-2\beta |J_2|}\right\}\sum_{\ell=2}^{\infty}(-1)^\ell e^{-\frac{\ell}{\xi}}\nonumber\\
    &\approx \left\{\frac{1}{4}\big[2 \ \sgn(J_2)-1\big]-\sgn(J_2)e^{-2\beta |J_2|}\right\}+\left\{\frac{1}{2}\big[1-\sgn(J_2)\big]+\big[\sgn(J_2)-1\big]e^{-2\beta |J_2|}\right\}\cdot\frac{1}{2}\nonumber\\
    &= \frac{\sgn(J_2)}{4}-\frac{1}{2}\big[1+\sgn(J_2)\big]e^{-2\beta |J_2|}+ O\left[e^{-4\beta |J_2|},e^{-2\beta (J_1-J_2)}\right],
    \label{eq:SumCorrelator}
\end{align}
where the correlation length $\xi$ is given by $\xi=-\left(\ln|g_{3}|\right)^{-1}$ and we have used the smallness
of the temperature, $T\ll |J_{2}|,\,2J_{1}-4J_{2}$, at the last step. 
Utilising Eqs.~\eqref{eq:sparsevacancy} and \eqref{eq:SumCorrelator}, we obtain the the susceptibility 
of the chain up to the first virial correction in the form
\begin{align}
    \chi (T) \approx\frac{N_{\text{vac}}}{2T}\big[1-\sgn(J_2)\big]+\frac{N_{\text{vac}}^2}{NT}\left\{\frac{\sgn(J_2)}{2}-\big[1+\sgn(J_2)\big]e^{-2\beta |J_2|}\right\}+ O\left[e^{-4\beta |J_2|},e^{-2(\beta J_1-J_2)}\right] +\chi_{\text{bulk}}(T),
\end{align}
where 
\begin{equation}
    \chi_{\text{bulk}}(T) = \chi_{0}+N_{\text{vac}}\,\sgn(J_{2})\,e^{-2\beta|J_{2}|} = \frac{N-b(T)N_{\text{vac}}}{N}\chi_{0}(T),
\end{equation}
with the vacancy size $b(T) = -\sgn(J_{2})\,e^{2\beta J_{1}-4\beta J_{2}-2\beta |J_{2}|}$. 
For Ising chains with antiferromagnetic NN and ferromagnetic NNN interactions, the first virial correction reduces the effective quasispin of the chain. In the case of antiferromagnetic NN and antiferromagnetic NNN interactions, the quasispins of isolated vacancies vanish, and the first virial correction introduces
leads to the $\propto N_\text{vac}^2/T$ quasispin behaviour of the susceptibility.



 \twocolumngrid
	

\end{document}